# Towards Alzheimer's Disease Progression Assessment: A Review of Machine Learning Methods

Zibin ZHAO

zzhaobz@connect.ust.hk

**Abstract:** Alzheimer's Disease (AD), as the most devastating neurodegenerative disease worldwide, has reached nearly 10 million new cases annually. Current technology provides unprecedented opportunities to study the progression and etiology of this disease with the advanced in imaging techniques. With the recent emergence of a society driven by big data and machine learning (ML), researchers have exerted considerable effort to summarize recent advances in ML-based AD diagnosis. Here, we outline some of the most prevalent and recent ML models for assessing the progression of AD and provide insights on the challenges, opportunities, and future directions that could be advantageous to future research in AD using ML.

## 1. Introduction

The disease associated with ageing problems places a significant sociological and psychological burden on the elderly due to the growing trend in older populations. Neurodegenerative brain disease, one of many illnesses linked to ageing, causes the population the most concern because of its severe effects and damage to the brain, which can result in symptoms like stroke and dementia or even be fatal (Sasaki, 1996). Neurodegenerative brain diseases include Parkinson's disease, ALS (amyotrophic lateral sclerosis), Alzheimer's disease (AD) and many others (*Brain Diseases*, n.d.). Due to its gradually and inherently increased risk with advancing age, Alzheimer's disease (AD), the most common cause of dementia in the world, receives much more attention than other brain diseases. AD is reported to have impacted on 5.8 million

Americans in 2020, and is estimated to reach 100 million by 2050 (Matthews et al., 2019), with the symptoms involves the gradual loss of memory, thought and language. Thus, requiring an improvement in the timely assessment and improvement of AD (S. Qiu et al., 2022). There have been recent efforts to identify the immune-relevant therapeutic target genes and monoclonal antibody-based medicines (Han et al., 2018; Tampi et al., 2021). Yet to date, there is still lack an efficient and resolve way for AD and its progression. The burden continues due to the aging effects with increased brain degeneration, increased number of patients, potential for error during the visual assessment of neuroimages and as-yet-unknown patterns and correlations among many biomarkers (Khojaste-Sarakhsi et al., 2022). Despite the fact that conventional diagnostic tools rely on the neurocognitive tests, assessing the levels of amyloid-β1-42 (Aβ42) or total tau protein and hyperphosphorylated tau (p-tau) (d'Abramo et al., 2020; Zeng et al., 2021), many studies have started to shed light on the AD assessment or even early diagnosis, especially with the boost from computational advancements. However, when introduced into the clinical setting without adequate fidelity and comprehension, many new emerging techniques continue to be difficult.

As a result of the advancement of cutting-edge imaging techniques, such as positron emission tomography (PET), and the availability of the Alzheimer's Disease Neuroimage Initiative dataset (*ADNI | Alzheimer's Disease Neuroimaging Initiative*, n.d.). Machine Learning (ML) and its subfield Deep Learning (DL) show promising results in assessing and providing insights in the early diagnosis of AD. Other open-source databases, such as the recent made public J-ADNI database (*Japanese ADNI Project*, n.d.), which includes longitudinal data in Japan, AIBL (*AIBL*, n.d.) and OASIS, also accelerates the ML society in classification tasks with the tremendous number of neuro-related data. Recently, several outstanding and comprehensive review papers on machine learning for AD have been published. In 2018, Jose et al. proposed a review of structural neuroimaging, specifically magnetic resonance imaging (MRI), for clinical prediction of a wide variety of brain disorders (Mateos-Pérez et al., 2018). Pellegrini et al. on the other hand (Pellegrini et al., 2018) screened 111 studies to provide a systematic review of multiple ML topics for identifying dementia from neuroimaging. Several other review studies discuss a review for timely diagnosis of AD (Fathi et al., 2022), and some mention ML in AD from omics, imaging and clinical data point of views to review the various prospective by ML to AD applications (Li et al., 2021). Conventionally, AD prognosis includes the classification problem of the different stages of NC, MCI (sMCI and pMCI) and AD, in order to provide an early pattern recognition for AD.

This review, written from the perspective of machine learning, provides insights for evaluating some contemporary methods for assisting in the early detection of AD. We present a summary of several significant ML models derived from a single or multiple data platforms, which may contribute to our understanding of the disease. Specifically, we divide our review topics into two main aspects: (i) supervised learning methods, (ii) unsupervised learning methods, encompassing both machine learning and its branch deep learning aspects. In addition to discussing some of the shortcomings and future opportunities, we aim to provide a comprehensive and easily-understood review and to suggest a number of challenges and opportunities that may arise from current practices.

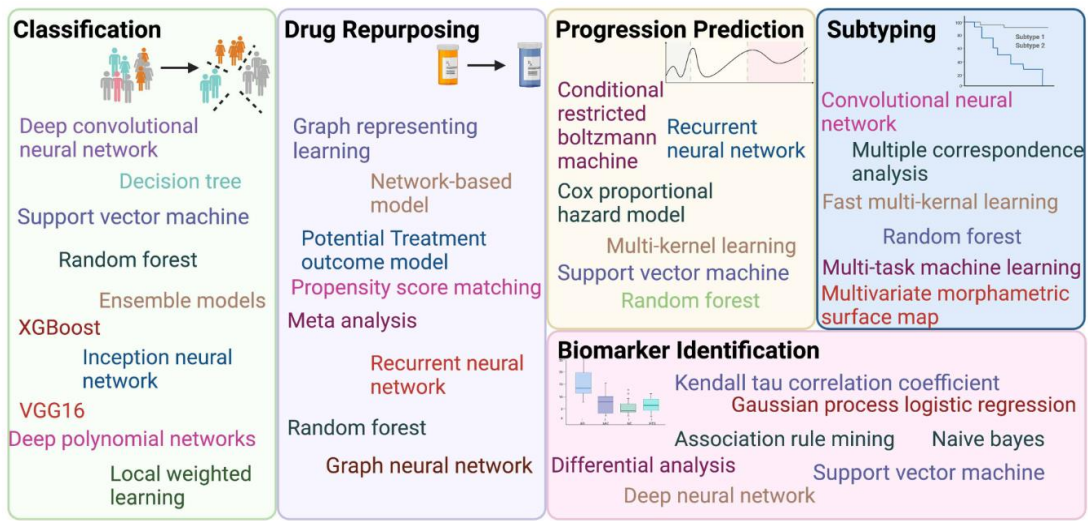

Figure 1. An overview machine learning methods applied to Alzheimer's disease research. (Li et al., 2021)

## 2. Search Strategy

Multiple databases and libraries, including Scopus, Web of Science, PubMed, and IEEE, were investigated to gain a deeper and more comprehensive understanding of the prominent performance of machine learning-based diagnosis for Alzheimer's Disease. The results revealed an upward trend in global interest in this field, particularly beginning in 2019.

## 3. Data Type

Prior to any model design, data selection is a crucial step in determining the model's architecture or type. For AD research, the most prevalent data types include omics, nominated target and drug repurposing, preclinical efficacy data, real-world patient data (neuropsychological tests and imaging), and knowledge repository. In ML model training, the most common types of imaging data for classification studies include positron emission tomography (PET), magnetic resonance imaging (MRI) which include structural MRI (sMRI) and functional MRI (fMRI) and electroencephalography

(EEG) (Li et al., 2021). In some cases, multi-modality images boost the diagnosis of AD using PET and MRI, where MRI provides a anatomical structural information and PET scan is able to provide the metabolic activities of the brain, thereby providing sufficient and useful information for AD prediction (Yang & Mohammed, n.d.).

## 3. Machine Learning

Machine learning (ML) is a branch of artificial intelligence (AI) that has long been proven to be a cost-effective tool for assisting in medical science particularly neuroengineering. It is governed by the laws of nature and employs the probability and statistics to generate task-specific models. ("What Is Machine Learning?," n.d.) However, the accuracy of AD classification tasks is highly dependent on the classification problem type. Normal Cognition (NC), Mild Cognitive Impairment (MCI), and Alzheimer's disease (AD) are the three common classification stages of AD researcher groups, whereas MCI can be further classified as stable MCI (sMCI) or progressive MCI (pMCI). The complete cycle of AD dementia could be depicted starting from NC to sMCI, then pMCI and finally AD. Current classification techniques distinguishing NC from MCI or NC from AD have reached an advanced stage. Despite this, it is observed that the classification of MCI versus AD is the most difficult due to the extensive overlap of features during model training (Tanveer et al., 2020). In the following sections, various model architectures and learning algorithms are discussed. Figure 1 depicts a general flowchart of the data collection, model development, and modal validation processes.

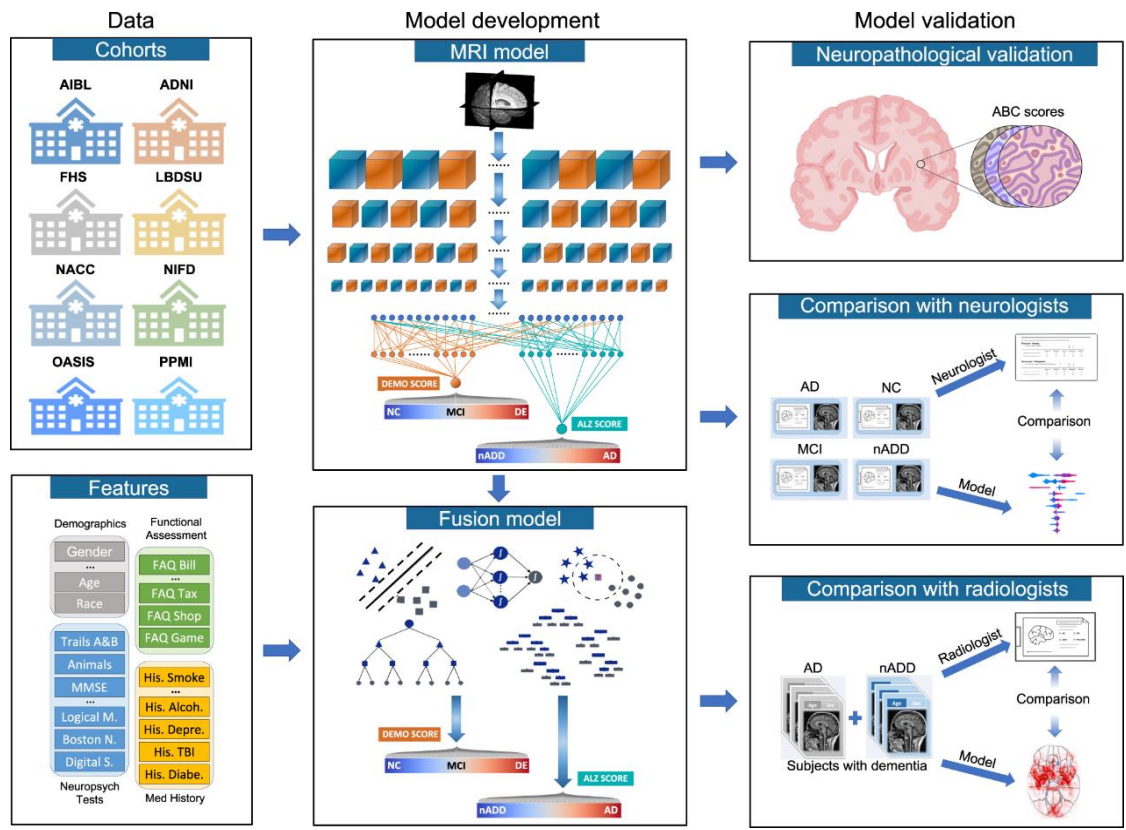

Figure 1. Modeling framework and overall strategy. (S. Qiu et al., 2022)

**4. Deep Learning**

As AI technology advances rapidly, deep learning (DL) becomes increasingly promising, particularly for imaging classification problems like computer vision or image segmentation (*What Is Perceptron: A Beginners Guide for Perceptron [Updated]*, n.d.). As a sub domain of ML, DL incorporates a model with a more complex architecture and deeper layers for more precise predictions. Several advances have been made in the classification of image data for the diagnosis of , which can be categorized mainly as supervised and unsupervised learning methods.

**5. Supervised Learning**

**5.1 DNN**

Beginning in the 1960s of the previous century, perceptron (*What Is Perceptron: A*

*Beginners Guide for Perceptron [Updated]*, n.d.) and neural networks with one input layer, one hidden layer, and one output layer began to appear. Deep Neural Networks (DNN) can be considered to include variant forms of CNNs and RNNs. However, in real-world applications, DNN typically employs multiple known structures, such as convolutional layers or LSTM units (explain in detail below).

**5.2 CNN**

Convolutional Neural Networks (CNN), the most predominant neural network architectures in image classification tasks, take advantage of the benefits of convolutional operation. The operation considers both spatial and temporal data characteristics and outputs classification results based on the weights (importance) of the input data features. Recently, DeepAD (Sarraf et al., 2016) employs a large number of training samples with both fMRI and MRI data in CNN architecture to address the issue of highly similar patterns in elderly. The final decision algorithm system achieves a 97.77% accuracy for fMRI and 100% for MRI images. In 2022, Heising and her team developed a 2D CNN based on LeNet-5 to accurately differentiate between AD and MCI, which share similar features and are difficult to distinguish using conventional ML (Heising & Angelopoulos, 2022).

**5.2 RNN**

Recurrent Neural Networks (RNN) are a class of deep learning that refers to an infinite impulse response with sequential input characteristics (*Recurrent Neural Network (RNN) Tutorial*, n.d.). RNNs are commonly deployed in speech recognition or natural language processing, with Long Short Term Memory (LSTM) or BiLSTM (bilayer LSTM) as their typical structure. The LSTM was adopted in (Aqeel et al., 2022) with neuropsychological measures (NM) and magnetic resonance imaging (MRI) as two biomarkers to be passed into the RNN to predict the patient with AD or MCI achieving an accuracy of 88.24%.

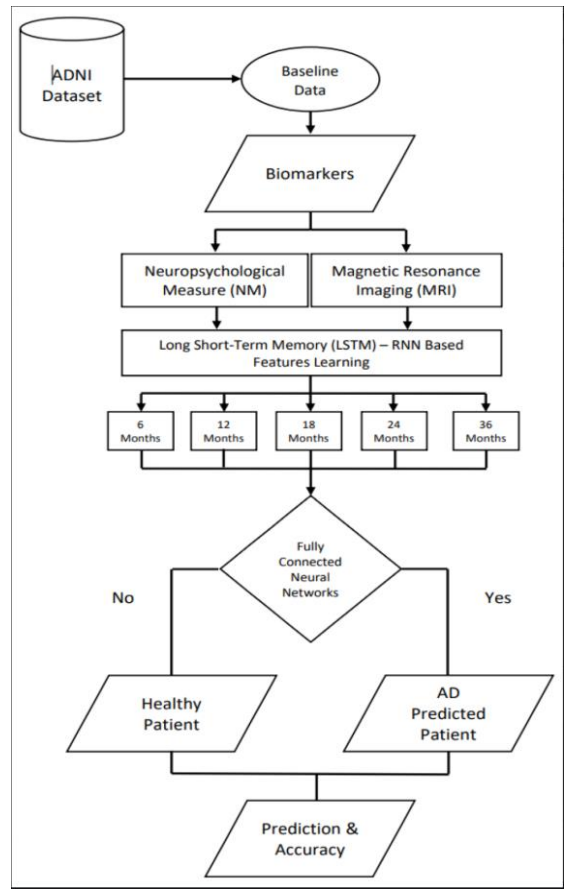

Figure 2. A typical RNN in the classification of healthy patient and AD predicted patient using NM and MRI biomarkers. (Aqeel et al., 2022)

In general, a rising number of researchers favor combining RNN and CNN to achieve a more accurate model based on sequential and special data. Qiu et al. proposed a graph-CNN-RNN network for AD prognosis on the ADNI dataset with an accuracy of 80% on AD conversion from 0 to 4 years prior to AD onset (A. Qiu et al., 2022). Nonetheless, both CNN and RNN requires a large amount of training data as well as optimized parameters and structure for improved prediction precision.

**5.4 DPN**

Deep Polynomial Network (DPN) is a variant of the DL algorithm that efficiently learns small feature representations from samples. DPN can be further categorized as Stacked DPN (S-DPN) for single modality or even multimodality as MM-S-DPN (Multimodality Stacked DPN) (Sharma & Mandal, 2022). Liu et al. in (Liu et al., 2015) suggested an MM-S-DPN for classification of multiple classes (AD vs. NC, MCI vs. NC) using PET and MR images for multimodal prediction.

## 6. Unsupervised learning

### 6.1 SVM

Support vector machine (SVM) was once the most popular ML models due to its ability to match input features to a very high dimensional output space and performs relatively accurate prediction. SVM differs from artificial neural network (ANN) with different optimization levels, global versus local (Bisgin et al., 2018). Some variants of SVM are twin SVM (TWSVM) and least-squares-based twin SVM (LSTSVM) to reduce the computational cost (Tanveer et al., 2020). In 2019, Gosztolyz et al. presented a linear SVM with 5-fold validation employing acoustic signal to predict the target of NC vs. MCI or vs. mAD (Gosztolya et al., 2019).

### 6.2 Auto-encoder (AE)

Encoder extracts latent features for representation, whereas decoder reconstructs information from input space (*Introduction to Autoencoders.*, n.d.). A variant could be the Stack autoencoder (SAE), in which multiple AEs are stacked to form a more compact feature extraction matrix. Liu S et al. were the first to address AD with DL, employing the SAE for early diagnosis of AD using PET and MRI images with three hidden layers (Liu et al., 2014).

## 7. Challenges and Opportunities

With the thrive in ML society, other topics related to AD with ML have recently been investigated, such as studying drug repurposing using transcriptome or network pharmacology (Paranjpe et al., 2019), studying subtyping of AD to identify the clinically homogenous group of patients, and even biomarker discovery for AD that alleviate the progression of AD. Additionally, there are multiple ML models that have not been mentioned due to their diminishing prevalence over time. Similar to DPN, the unsupervised model Deep Boltzmann Machine (DBM) or restricted BM (RBM) has emerged as a technique that generates the probability distribution of the input dataset. There is, however, a trade-off between model complexity and accuracy. CNN and its variants continue to be the predominant DL model at present.

Based on the aforementioned models, both ML and DL methods exhibit an exceptional capacity for integrating multiple data types and platforms for AD classification. However, obstacles remain from the following perspectives: 1) the reliability of AI technology in explaining the prediction results; 2) the implementation of patient data in the real world for accurate identification; and 3) the advancements with multimodal ML or DL. In ML society, there has always been such a "black box" that the intermediate steps or hidden layers of the architecture have not been adequately explained. Consequently, imposing a problem with less or even unreliable data for clinical applications. Scientists have been working on trustworthy AI for a long time in order to provide a better understanding of the prediction results, such as using gradient class activation map (GRAD-CAM) to emphasize the important features in an image that the model used to calculate the prediction score (Selvaraju et al., 2020).

Here, we encourage a greater emphasis on the model's explanatory mechanism, embedded with both ante-hoc and post-hoc explanations to increase model fidelity. In addition, a benchmarking should be standardized for publishing such clinically relevant articles by revealing the model's transparency. As per observation, the use of the Transparent Reporting of a multivariable prediction model for Individual Prognosis or Diagnosis (TRIPOD) (Collins et al., 2015) should be increased as none of the published models provide a standard benchmark. More in more, there is significant room for improvement in the discrimination between MCI vs. NC and MCI vs. AD. Hence, the research group should be more cognizant of the significance of utilizing multimodal, as the majority of the models only consider imaging data. Nonetheless, omics data types such as genomics, metabolomics, and epigenetics may provide more accurate pathways for predicting alterable gene or metabolite targets, which could be a promising direction for ML and DL research in the near future (Li et al., 2021).

## 8. Conclusion

The primary objective of this review is to predict the progression of Alzheimer's disease using machine learning and other discovery techniques gleaned from various research databases. In the past decade, the field has rapidly embraced the power of machine learning for sophisticated data analysis and integration. In addition, there is a growing trend towards applying techniques of deep learning to exploit the massive and complex data in AD research.